\documentclass[nofootinbib,notitlepage,superscriptaddress,10pt,aps,pra,twocolumn]{revtex4-1}

\usepackage[utf8x]{inputenc}
\usepackage{amsmath}
\usepackage{amssymb}
\usepackage{bbm}
\usepackage{graphicx}
\newcommand{\be}{\begin{equation}}
\newcommand{\ee}{\end{equation}}
\newcommand{\beq}{\begin{eqnarray}}
\newcommand{\eeq}{\end{eqnarray}}
\newcommand{\bea}{\begin{eqnarray}}
\newcommand{\eea}{\end{eqnarray}}
\begin{document}
\title{Lorentz invariant deformations of momentum space}

\author{Valerio ASTUTI}
\affiliation{Dipartimento di Fisica, Universit\`a di Roma ``La Sapienza", P.le A. Moro 2, 00185 Roma, EU}
\affiliation{INFN, Sez.~Roma1, P.le A. Moro 2, 00185 Roma, EU}
\affiliation{Perimeter Institute, 31 Caroline St N, Waterloo ON, N2L 2Y5, Canada}

\author{Laurent FREIDEL}
\affiliation{Perimeter Institute, 31 Caroline St N, Waterloo ON, N2L 2Y5, Canada}

\begin{abstract}
In relative locality theories the geometric properties of phase space depart from the standard ones given by the fact that spaces of momenta are linear fibers over a spacetime base manifold. 
In particular here it is assumed that the momentum space is non linear and can therefore carry non trivial metric and composition law. We classify to second order all possible such deformations that preserve Lorentz invariance.
We show that such deformations still exists after quotienting out by diffeomorphisms only if the non linear addition is non associative.\end{abstract}

\maketitle

\section{Introduction}
The search for a quantum gravity theory gave us many hints that what we know about Lorentz symmetry has to be changed in order to make room for an invariant scale of distance and energy, the so called Planck scale.
There are heuristic arguments leading to the existence of an absolute, minimum scale of distance, $L_P=\sqrt{\frac{G \hbar}{c^3}}$. This scale should be the same in every reference frame in order not to break the Lorentz symmetry between observers. 
We know however that the notion of an absolute length scale is incompatible with the standard Lorentz group of symmetries. Trying to incorporate such a lenght scale in a symmetry group led  to the  introduction of  DSR (Doubly Special Relativity) theories, in which the Lorentz group is deformed in a more general class of objects called quantum groups \cite{DSR1, DSR2, DSR3, DSR4}.
In recent times DSR theories have been linked to the concept of \emph{relative locality} \cite{reloc, Qrel1, Qrel2}. This concept is the realization that every inference we make about spacetime derives from observation of particles properties, like their energy and momentum, through their interaction. In this picture  spacetime is a derived quantity resulting from the interactions of probes.
This perspective is in agreement with particle physics  experiments, in which the only measured quantities are momenta and timings, from which positions are subsequently derived.
 This is somewhat constrasting with the standard  view of mechanics, where momenta of particles are entities derived from spacetime measurements, and correspondingly the momentum space is defined trough spacetime concepts. 
 The reconciliation of this tension lies in the fact that the proper geometrical setting to describe the physics of localisation is phase space itself \cite{FLM}. One of the key assumption of relative locality is that in a particular high energy limit of gravity 
 we can focus entirely on the geometry of momentum space \cite{reloc}.
 
 A particularly striking consequence of this realization is that spacetime itself, as an absolute object shared between every observer, ceases to be a necessary assumption. In this limit momenta are the most basic entities and building the framework of mechanics around this principle we can allow for nontrivial geometries in momentum space - which becomes a nonlinear manifold - and obtain spacetime as attached fibers on every momentum space point.  
In particular the absoluteness of spacetime is equivalent to the flatness of the momentum manifold, feature that is not inevitable when we take into consideration the possible deformations implied by the Planck scale. 

In general we can have a momentum space which has nontrivial metric, curvature, and torsion. The symmetry transformations of this space are no more elements of the Lorentz group, but of some deformation of it. 
In this work we present a full analysis to second order in the deformation parameter of the most general non linear structure which admits a fully compatible action of the Lorentz group.
The study of the generalization of the properties of momentum space manifolds and in particular of Lorentz invariant composition laws have been initiated only recently in \cite{Palm, BF}.

\section{Phase space bundle structure}
The usual bundle structure of the phase space is the following: we start from the spacetime manifold $M$, and consider associated to each point $x$ of it a cotangent space, the momentum space $T^*_x M$.
In the usual flat, continuous spacetime, the structure of this cotangent bundle is very simple. We have a global trivialization which allows us to identify the cotangent spaces relative to different points, and - in the flat case - the possibility to identify the cotangent space with the manifold itself. So we can consider as phase space the product space $\mathbb{R}^{2n}$.

From this very simple, trivial case, we depart in a direction different from the usual one. Instead of giving room for a curvature of the spacetime manifold $M$, we revert the relation between the two sectors of the bundle and consider the momentum space, denoted from now on $P$, as the base manifold. In this way, in the general case, we have no more a manifold $M$ as spacetime, but a cotangent bundle $T^*_p P$. 

On the other hand now the manifold $P$ is no longer forced to be a trivial one but can have a curvature, a torsion, and a nontrivial metric. In addition, being a non-flat manifold, we are not allowed to identify the tangent spaces with the manifold itself. We have to distinguish points of the manifold, identified by their coordinates $p_{\mu}$, and points in its tangent spaces, identified as usual as $\{v_{\mu}, w_{\mu}, ...\}$.

Of course at zeroth order approximation the manifold $P$ will be flat, torsionless and with trivial metric (so we will be able to make the identification with the spacetime cotangent spaces). But we are looking for possible corrections to this zeroth order approximation, allowing for deformations of the symmetry group, the metric and the composition law of momenta.
We have, perturbing around the flat-manifold approximation, the possible deformations respectively for the symmetry group, the metric and the composition law:
\begin{equation}
\Lambda(p)_{\mu} = \Lambda_{\mu}^{\nu}p_{\nu} +  \frac{\Delta_{\mu}^{\nu\rho}}{2M} p_{\nu}p_{\rho} + 
O\left(\frac{p^2}{M^2}\right)
\end{equation}
\begin{equation}
g_{p}(v,w) = \left(\eta^{\mu \rho} + \frac1M G^{\nu, \mu \rho}p_{\nu} + O\left(\frac{p^2}{M^2}\right)\right) v_{\mu}w_{\rho} 
\end{equation}
\begin{equation}
(p\oplus q)_{\mu} = (p + q)_{\mu} + \ell S_{\mu}^{\nu\rho} p_{\nu}q_{\rho} + \cdots+ O\left(\frac{p^2}{M^2}\right)
\end{equation}
where $M$ is a  mass  scale linked to the Planck mass, and the dots in the last equation are terms proportional to $pp$ and $qq$ products.

In the following we will denote with $p_{\mu}$ momenta expressed in units of $M$, for the sake of notation. With this notation in every power series in the momentum variable each additional term will introduce a further suppression factor of $M$.

The only constraint we put on the deformation of this momentum-based phase space is the validity of the symmetry group as an isometry for the new metric, and as an homomorphism of the new composition law. We look for the most general cotangent bundle compatible with a possibly deformed Lorentz symmetry. To be more precise, we ask that the metric is left unchanged by a Lorentz transformation:
\begin{equation}
g_{\Lambda(p)}(d\Lambda(v),d\Lambda(w)) = g_{p}(v,w)
\end{equation}
and the same transformation to be an homomorphism of the composition law:
\begin{equation}
\Lambda(p\oplus q)=\Lambda(p)\oplus \Lambda(q)
\end{equation}
It is important to note that this last condition excludes the so-called $\kappa$-Poincar\'e composition law.
The $\kappa$-Poincar\'e composition law  is an addition law on deSitter space, which {\it is not} preserved by Lorentz transformations in the sense just described (see e.g \cite{Kowalski}).

\section{Triviality of the symmetry group deformation}
We start asking for all the possible deformations of the Lorentz group in this setting. The answer to this question is pretty simple: we have none in the Lie algebras category \cite{Mendez, JV, BF}. 
It is easy to prove that for the Lorentz algebra (and all semi-simple algebras) all possible deformations of the algebra structure amount to a redefinition of its generators, so it can be trivialized by a change of coordinates in the generators space (or a change of parameters).
To prove this fact we need to define a little of cohomology theory (we essentially go through the same derivation as in \cite{Mendez}). Given a Lie algebra $L$: 
\begin{equation}
[A,B]=if_{AB}^C C
\end{equation} 
and its adjoint representation 
\begin{equation}
\rho_A(B)=[A,B]
\end{equation}
with $A, B, C$ in some vector space $V$, we define an \emph{n-cochain} $\gamma$ in the adjoint representation as a multilinear skew-symmetric map of $n$ copies of $V$ into itself:
\begin{equation}
\gamma : V \times V \times ... \times V \to V
\end{equation}
Obviously the set of all n-cochains forms a vector space $C^n(\rho, V)$.
The \emph{coboundary operator} $d : C^n(\rho, V) \to C^{n+1}(\rho, V)$ is the linear map:
\begin{equation}
\begin{split}
d \phi (A_1, ... , A_{n+1}) = \sum_{i=1}^{n+1}(-)^{i-1}\rho_{A_i}\left(\phi(A_1,...,A_{n+1})\right) + \\
+ \sum_{1\leq i < j \leq n+1}(-)^{i+j}\phi\left([A_i,A_j],A_1,...,A_{n+1}\right)
\end{split}
\end{equation}

where in the first line we don't have $A_i$ among the arguments of $\phi$, and in the second line we don't have neither $A_i$ nor $A_j$ ($\phi \in C^n(\rho, V)$ is an n-dimensional cochain). 
An n-cochain $\phi$ is called an \emph{n-cocycle} whenever we have $d\phi = 0$. The vector space of all n-cocycles is denoted $Z^n(\rho)$.

An n-cochain $\phi$ is called an \emph{n-coboundary} if it is in the image of the coboundary operator, $\phi \in d\left(C^{n-1}(\rho, V)\right)$. The vector space of all n-coboundaries is denoted $B^n(\rho)$.

For our scopes we will consider the quotient space $H^n(\rho) = Z^n(\rho)/B^n(\rho)$, called the \emph{n-cohomology group}. While from the definitions given above we have $d^2\phi = 0$ for every $\phi$, as in the standard de Rham cohomology not all cocycles are coboundaries, and the cohomology group may present a nontrivial structure. 

The next step we need is the definition of a Lie algebra deformation. Given the Lie algebra $L_0$:
\begin{equation}
[A,B]=if_{AB}^C C
\end{equation}
a one-parameter deformation $L_t$ of it is defined as:
\begin{equation}
[A,B]_t = [A,B] + \sum_{i=1}^{\infty}\phi(A,B)_i t^i
\end{equation}
where $\phi(A,B)_i$ are 2-cochains.

We want the new algebra to be a Lie algebra, so the deformed commutators must comply with the Jacobi identities for all $t$:
\begin{equation}
[[A,B]_t,C]_t + [[B,C]_t,A]_t  = [[A,C]_t,B]_t \quad \forall t
\end{equation} 
Differentiating this condition with respect to $t$ and setting $t=0$ we obtain the condition:
\begin{equation}
d\phi_1(A,B,C)=0
\end{equation}
i.e. $\phi_1$ must be a 2-cocycle. 

This is why cohomology theory is so important for Lie algebra deformations: a fundamental result in cohomology theory of Lie algebras \cite{cheveil} is that any semi-simple Lie algebra (like the Lorentz one) has trivial second cohomology group, $H^2(\rho)=\{0\}$. This implies that $\phi_1$ must be a coboundary:
\begin{equation}
\phi_1=d\phi_0
\end{equation}
for some 1-cochain $\phi_0$. 
Let us now consider the new algebra $[A,B]_t'$ obtained through the map $F_t=e^{-t\phi_0}$:
\begin{equation}
[A,B]_t' = F_t^{-1}\left([F_tA,F_tB]_t\right)
\end{equation}
It is easy to see that this transformed algebra has a trivial first 2-cochain:
\begin{equation}
\begin{split}
\phi_1'(A,B) = \phi_1(A,B) + \phi_0([A,B]) +\\
- [\phi_0(A),B] - [A,\phi_0(B)] = 0 \label{zero}
\end{split}
\end{equation}
where the last equality comes from the definition of $\phi_0$: $\phi_1 = d\phi_0$.
From the equation \eqref{zero} we conclude that the deformation of the new algebra starts from the second order term:
\begin{equation}
[A,B]'_t = [A,B] + \sum_{i=2}^{\infty}\phi'_i(A,B)t^i
\end{equation}
But now $\phi'_2(A,B)$ must be a 2-cocycle, so we can iterate the process, to show that every deformation of a semi-simple Lie algebra can be gauged away by a linear transformation.
So any deformation of the Lorentz algebra (that is restricted to the Lie algebra category) can be reabsorbed in a redefinition of the generators. The Lorentz algebra, along with every semi-simple algebra, is called \emph{stable} under deformations\footnote{What we exposed here is the so-called Nijenhuis and Richardson theorem.}.

\section{Deformations of the metric}
Given the stability of the Lorentz group under deformations we can assume that we work in coordinates where the Lorentz group action assumes the  standard form, and go on to look for all possible deformations of the metric. 
In doing so we work in perturbation theory around $p=0$, where we know the metric has to be the standard one:
\begin{equation}
g_{p}^{\mu \rho} = \left(\eta^{\mu \rho} + G^{\nu, \mu \rho}p_{\nu} + O\left(p^2 \right)\right)
\end{equation}
In order for the metric to be Lorentz invariant all the tensors that enter this power series have to  be Lorentz covariant\footnote{We are in coordinates in which the Lorentz group is not deformed, so $d\Lambda = \Lambda$.}:
\begin{equation}
g_p(v,w) = g_{\Lambda p}(\Lambda v,\Lambda w) \Rightarrow G^{\alpha, \beta \gamma} = G^{\nu, \mu \rho} \Lambda_{\nu}^{\alpha}\Lambda_{\mu}^{\beta}\Lambda_{\rho}^{\gamma} 
\end{equation} 
So we are limited in our choice to functions of the undeformed metric $\eta^{\mu \nu}$, or the totally antisymmetric tensor $\epsilon^{\alpha \beta \gamma \delta}$. Both of them has only an even number of indices, so we cannot build a Lorentz covariant tensor with an uneven number of indices, and this rules out all the uneven orders in the power series. 
For the even orders, we have another strong constraint given by symmetry with respect to permutation of the indices. This condition rules out the antisymmetric tensor $\epsilon^{\alpha \beta \gamma \delta}$. The metric should be symmetric for the exchange of the two ``tangent" indices, and every coefficient of the series is symmetric under the exchange of ``coordinate'' indices. For example, at the second order of the series $G_2^{\mu \nu, \sigma \rho}p_{\sigma}p_{\rho}$, to build the tensor $G_2^{\mu \nu, \sigma \rho}$ as a function of the undeformed metric with the right symmetries we have only two independent choices, so the most general form for this term will be:
\begin{equation}
G_2^{\mu \nu, \sigma \rho} = A_2 \eta^{\mu \nu}\eta^{\sigma \rho} + B_2 \left( \eta^{\mu \rho}\eta^{\sigma \nu} + \eta^{\mu \sigma}\eta^{\nu \rho} \right) 
\end{equation}
Similarly, for each higher order term, we have that the space of possible tensors with the right symmetries is two-dimensional, one independent component being the one in which the two tangent indices are coupled in the same tensor and the coordinate ones are symmetrized in the others, and the other one with the totally symmetric combination of tangent and coordinates indices:
\begin{equation}
\begin{split}
G_{2n}^{\mu \nu, \sigma \rho \xi \eta ...} = A_{2n} \eta^{\mu \nu}\left[Sym\left( \eta^{\sigma \rho}\eta^{\xi \eta}...\right)\right] + \\ + B_{2n} \left[Sym\left( \eta^{\mu \sigma}\eta^{\nu \rho}\eta^{\xi \eta}...\right)\right] 
\end{split}
\end{equation}   
When this particular combination of indices is contracted with the momentum and the tangent vectors we obtain:
\begin{equation}
\begin{split}
G_{2n}^{\mu \nu, \sigma \rho \xi \eta ...}v_{\mu}w_{\nu}p_{\sigma}p_{\rho}p_{\xi}p_{\eta}...= \\ = A_{2n} p^{2n} v\cdot w  + B_{2n} p^{2(n-1)} (p\cdot v)(p\cdot w)
\end{split}
\end{equation}
Of this coefficients we know the lowest order ones, those of the undeformed metric:
\begin{equation}
A_{0}=1 \qquad B_{0}=0
\end{equation}
Now we can rewrite the full metric as a power series:
\begin{equation}
\begin{split}
g_p(v,w) = \sum_{n=0}^{\infty} p^{2n} \left[ A_{2n}  v\cdot w  + B_{2(n+1)}  (p\cdot v)(p\cdot w) \right]
\end{split}
\end{equation}
At this point we can resum the series (the momenta are expressed in adimensional units, as multiples of the mass $M$, so we have $p^2 \ll 1$) to obtain the general form:	
\begin{equation}
g_p(v,w) = A(p^2) v\cdot w  + B(p^2) (p\cdot v)(p\cdot w)
\label{metdef}
\end{equation}
with $A$ and $B$ analytic functions with the constraints:
\begin{equation}
\lim_{p^2\to 0} A(p^2)=1 \quad, \quad  \lim_{p^2\to 0} B(p^2)=0
\end{equation}
This metric cannot be reduced to the standard one using a diffeomorphism, unless the two functions $A(p^2)$ and $B(p^2)$ satisfy a specific condition. To prove this, we write the metric as a function of the frame field $e_{\alpha}^{\beta}(p)$:
\begin{equation}
g^{\mu \nu}_{p} = \eta^{\alpha \beta}e_{\alpha}^{\mu}(p)e_{\beta}^{\nu}(p) 
\end{equation}
In order to have a metric like the one in (\ref{metdef}), the frame field must be of the form:
\begin{equation}
e_{\alpha}^{\mu}(p) = f(p^2)\delta_{\alpha}^{\mu} + g(p^2)p^{\mu}p_{\alpha}
\end{equation} 
with 
\begin{equation}
f(p^2)=\sqrt{A} \quad , \quad g(p^{2}) = \frac{1}{p^2}\left(\sqrt{A+{B}p^2}-\sqrt{A}\right) 
\end{equation}
Now for the frame field to be a diffeomorphism, $e_{\alpha}^{\mu}=\frac{\partial p'_{\alpha}}{\partial p_{\mu}}$, it must comply with the condition: 
\begin{equation}
\frac{\partial e_{\alpha}^{\mu}(p)}{\partial p_{\nu}} = \frac{\partial e_{\alpha}^{\nu}(p)}{\partial p_{\mu}}.
\end{equation}
This condition is satisfied if and only if:
\begin{equation}
g(p^2)=2\frac{\partial f(p^2)}{\partial p^2}.
\end{equation}
When the frame field satisfies this condition we can trivialize the metric with the diffeomorphism:
\begin{equation}
p'_{\mu} = f(p^2)p_{\mu}.
\end{equation}
In all other cases the transformation cannot be reabsorbed in a change of coordinates.

\section{Deformations of the composition law}
We can make a similar analysis for the coefficients in the composition law power series:
\begin{equation}
(p\oplus q)_{\mu} = (p + q)_{\mu} +  S_{\mu}^{\nu\rho} p_{\nu}q_{\rho} + ... + O\left(p^2 \right)
\end{equation}
The coefficients $S_{\mu}^{\nu\rho ...}$ must be Lorentz covariant, so we can rule out all the ones with an uneven number of indices for the reasons exposed above, and the totally antisymmetric tensor $\epsilon^{\alpha \beta \gamma \delta}$ (it has four free antisymmetric indices, but we have only two momenta in the composition law): we are left again with products of the undeformed metric tensor.
For example, for the lowest order terms we have the only possibility:
\begin{equation}
\begin{split}
(p\oplus q)_{\mu} = (p+q)_{\mu} + p_{\mu}(Ap^2 + B p\cdot q + Cq^2) + \\ + q_{\mu} (Dp^2 + E p\cdot q + F q^2) + ...
\end{split}
\end{equation}
For the higher orders it is easy to replicate the derivation given for the metric, leading to the expected solution that we are allowed to modify the composition law between two arbitrary momenta by coefficients given by analytic scalar functions, having value close to $1$ for small momenta:
\begin{equation}
(p\oplus q)_{\mu} = S_1(p^2,p\cdot q,q^2)p_{\mu} + S_2(q^2,p\cdot q,p^2)q_{\mu}
\end{equation}
\begin{equation}
\lim_{p,q\to 0} S_i(p^2,p\cdot q,q^2) = 1 \qquad i=1,2
\end{equation}
With the hypothesis of $q=0$ being the identity element of this addition, we obtain:
\begin{equation}
S_{i}(p^2,0,0)=1
\end{equation}
At the lowest order in the $p,q$ expansion these functions can therefore be given by 
\bea
S_{i}(p^2,p\cdot q,q^2) = 1+ A_{i}q^{2}+B_{i}p\cdot q + o(p^{2})
\eea
that is each coefficient depends only on two of the three arguments at this order.
The addition transforms covariantly under diffeomorphism. Given a diffeomorphism $F$ we get the transformed addition as 
\be
(p\oplus' q)_{\mu} \equiv F\left(\left[F^{-1}(p)\oplus F^{-1}(q)\right]\right)_{\mu}
\ee
At the lowest non-trivial order in the deformation if one takes $ F(p)_{\mu} = (1 - \alpha p^{2}) p_{\mu} + o(p^{3})$, the transformed addition is related to the old one by the transformation
\bea
A_{i}'= A_{i}-\alpha,\qquad B_{i}'= B_{i}-2 \alpha.
\eea 
Let us also note that we can change the values of the coefficients by a rescaling of the momenta:
under  $ p \to  p/\lambda$ the coefficients transforms as $ A_{i}\to \lambda^{2} A_{i}$, $B_{i}\to \lambda^{2}B_{i}$.
This means that up to diffeomorphisms and rescaling we have at lowest non-trivial order a two parameters family of Lorentz invariant composition laws.

When one of the two momenta is infinitesimal this composition law should reduce to the sum of the finite momentum plus the infinitesimal one, rotated by a transport to the point coordinatized by the first momentum:
\begin{eqnarray}
(p\oplus \delta q)_{\mu} &\approx & S_1(p^2 ,p\cdot \delta q,0)p_{\mu} + S_2(0 ,0 ,p^2)\delta q_{\mu}  \quad \\
&\approx & p_{\mu} + U_{\mu}^{\rho}(p) \delta q_{\rho} \nonumber
\end{eqnarray}
with 
\begin{equation}
U_{\mu}^{\nu}(p)=\left(\frac{\partial S_{1}}{\partial p\cdot q} \right)_{q=0} p_{\mu}p^{\nu} + S_2(0 ,0,p^{2})\delta_{\mu}^{\nu}
\end{equation}
which represents the left transport operator.

At the lowest order it is given by:
\be
U_{\mu}^{\rho}(p)= A_{2} p^{2}\delta_{\mu}^{\nu} + B_{1}p_{\mu}p^{\nu}+ o(p^{2}).
\ee
It is now interesting to compute the curvature tensor at the origin.
 We use the definition given in \cite{reloc}:
\begin{equation}
\Lambda(0)^{bc a}{}_{\mu } = \left.\frac{\partial}{\partial k_{a}}\frac{\partial}{\partial p_{b}}\frac{\partial}{\partial q_{c}}\left((p\oplus q)\oplus k - p\oplus(q\oplus k) \right)_{\mu}\right|_{p,q,k=0}, \nonumber
\end{equation}
which measure the lack of associativity at the origin.

The definition given in \cite{reloc} for the curvature is $R(0)^{bca}{}_{\mu }=2 \Lambda(0)^{[bc] a}{}_{\mu}$.
From this definition we obtain a curvature of the form:
\begin{equation}
R(0)^{bca}{}_{\mu }=\left( \delta_{\mu}^{c}\eta^{ab}-\delta_{\mu}^{b}\eta^{ac}\right) \left( 2 \frac{\partial S_{1}}{\partial q^2}- \frac{\partial S_{2}}{\partial p\cdot q} \right)(0,0,0).
\end{equation}
More generally the associativity $\Lambda^{bc a}{}_{\mu}$ at the origin is measured by 
\be
\eta^{ab}\delta^{c}_{\mu}(B_{1}-B_{2})  +\eta^{ac}\delta^{b}_{\mu}(B_{1}-2A_{1}) -  \eta^{bc}\delta^{a}_{\mu}(B_{2}-2A_{2}).
\ee
Demanding this to vanish implies that the addition is given by the linear addition up to a diffeomorphisms.
This implies that Lorentz invariance plus associativity excludes at this order any possible deformation. A non trivial lorentz invariant addition is necessarily non associative.

\section{Inverses}
Given the general structure for a Lorentz covariant momentum manifold we can ask for additional structures, and investigate the consequences in terms of metric and composition law of these additional constraints. 
To do so we work at the second order in perturbation expansion, which is  interesting in terms of physical consequences.

Given the general form of the composition law it is easy to derive the form of the left and right inverse of the operation:
\begin{equation}
\begin{split}
\ominus_{L} p \oplus p 
= 0,\qquad p \oplus(\ominus_{R} p) = 0.
\end{split}
\end{equation}
These are given by 
\bea
 \ominus_L p = - \frac{S_2( p^2, \ominus_L p \cdot  p, \ominus_L p^2 )}{S_1(\ominus_L p^2, \ominus_L p \cdot p, p^2)} p,  \\
 \ominus_R p = - \frac{S_1(p^2, p\cdot \ominus_R p, \ominus_R p^2)}{S_2(\ominus_R p^2,  p\cdot \ominus_R p ,p^2)} p.
\eea
The expressions above can be perturbatively solved for $\ominus p$. We obtain:
\begin{equation}
\ominus_L p = - \left[ 1 + \left(A_2 -A_{1} +  B_1  - B_2 \right) p^2 \right] p  
\end{equation}
\begin{equation}
\ominus_R p = - \left[ 1 - \left(A_2 - A_{1} +  B_1  - B_2 \right) p^2 \right] p  
\end{equation}
For the two inverses to be the same we need to satisfy the condition:
\begin{equation}
(A_1 - B_1)=(A_2 - B_2) \label{eq:invsim}
\end{equation}
which implies on the other hand that the inverses are undeformed. 
The set of Lorentz invariant composition laws which have the same left and right inverses up to diffeomorphism form a two dimensional family. We can express this addition family in a standard form after performing an infinitesimal diffeomorphism
\be
F(p)_{\mu} = (1-(B_{1}-A_{1})p^{2}) p_{\mu} +\cdots
\ee
The most general addition rule with same left and right inverse is to second order given by
\be
(p\oplus q)_\mu
= \left(1+ \beta_{1}(q^{2}+ p\!\cdot\!q) \right)p_{\mu} +
\left(1+ \beta_{2}(p^{2}+ p\!\cdot\!q)\right) q_{\mu},\ee
where $\beta_{i} \equiv 2A_{i}-B_{i}$.

We obtain even stronger constraints from the condition of ``left invertibility'':
\begin{equation}
\ominus p \oplus (p \oplus q) = q \qquad \forall p,q \in P \label{eq:cominv}
\end{equation} 
This condition implies on the lowest order coefficients the constraints:
\begin{eqnarray}
2A_2-B_{2} &=& 0 \\
(A_1 - B_1)  &=& (A_2 - B_2)  \nonumber
\end{eqnarray}
So the condition of left invertibility implies the symmetry between left and right inverses. 
We see that the demand of left invertibility implies that $\beta_{2}=0$ so the left invertible composition laws 
are all given after a diffeomorphism  by
\begin{equation}
p\oplus q = \left[1 + \beta_{1} ( q^2 +  p\cdot q)\right] p + q
\end{equation}
This  correspond exactly to the expansion of the  composition law of the kind presented in \cite{BF}.
In this work a general left invertible and Lorentz invariant composition law was introduced and studied.

Moreover, if we also demand the right invertibility we have the additional condition 
\be
2A_{1} - B_{1}= 0,
\ee
which implies that the composition law is associative, hence trivial.
The demand of Lorentz invariance plus right and left invertibility 
thus implies that the composition law is trivial at the lowest non-trivial order.

Another interesting property of the inverse we can investigate is that of it being a \emph{morphism} or an \emph{antimorphism}, expressions meaning, respectively:
\begin{equation}
\ominus (p\oplus q) = (\ominus p) \oplus (\ominus q)
\end{equation}
or
\begin{equation}
\ominus (p\oplus q) = (\ominus q) \oplus (\ominus p)
\end{equation}

The condition for the composition rule to be a morphism brings simply to a repetition of the condition of symmetry between left and right inverse \eqref{eq:invsim}, so 
the morphism property is implied by the symmetry between left and right inverses.
 The antimorphism condition, on the other hand, requires a complete symmetry of the composition law\footnote{This obviously implies the morphism condition.}:
\begin{equation}
A_1 = A_2 \qquad B_1=B_2
\end{equation}
So, for example, it is impossible to have a non trivial composition law with the property of left invertibility \eqref{eq:cominv} and which is an antimorphism.

\section{Metric compatibility}
The first Lorentz invariant example of momentum manifold has been presented in \cite{BF}. In this example composition law and metric were demanded to be compatible.
That is, it was required that  the metric is left invariant by the transformation generated by 
a left addition
\begin{equation}
g_{L_q(p)}\left(d_{p}L_q(v),d_{p}L_q(w)\right) = g_{p}(v,w) \quad , \quad L_q(p) = q\oplus p 
\end{equation} 
where $ d_{p}L_q$ denotes the differential at $p$ of $L_{q}$.
Having at hand the general form for metric and composition rule we can check if the condition of metric compatibility is enough to recover the manifold the authors found in \cite{BF}, or if there are other options realizing this condition. They find for the metric and composition law the expressions:
\begin{equation}
g_p^{\mu \nu} = \eta^{\mu \nu} + p^{\mu}p^{\nu} + o(p^2)
\end{equation}
\begin{equation}
(p\oplus q)_{\mu} = \left[ 1 - \frac{1}{2}\left(p\cdot q + q^2\right)\right]p_{\mu} + q_{\mu} +o(p^3)
\end{equation}

Expressing in our setting the metric and composition law as power series we have:
\begin{equation}
g_p^{\mu \nu} = (1 + Ap^2) \eta^{\mu \nu} + B p^{\mu}p^{\nu} + o(p^2)
\end{equation}
\begin{equation}
(p\oplus q)_{\mu} = (1 + A_1 q^2 + B_1p\cdot q )p_{\mu} + (1 + A_2 p^2 + B_2 p\cdot q)q_{\mu} + o(p^3)
\end{equation}
We now look at this order to the most general solution of the metricity condition, to obtain that 
\begin{equation}
\begin{split}
(p\oplus q)_{\mu} = \left(1 + \frac{(A-B)}{2} q^2 - \frac{B}{2} p\cdot q \right)p_{\mu} + \\+ \left(1 -\frac{A}{2} p^2 - A p\cdot q \right)q_{\mu} + o(p^3)
\end{split}
\end{equation}
Though the previous expression looks different from the one found in \cite{BF}, we can always perform a diffeomorphism on our variables; in particular, choosing 
\begin{equation}
p'(p)_{\mu} = \left(1 + \frac{A}{2}p^2 \right)p_{\mu}
\end{equation} 
we obtain:
\begin{equation}
g_{p'}^{\mu \nu} = \eta^{\mu \nu} + \left(B - 2A\right) p'^{\mu}p'^{\nu} + o(p'^2)
\end{equation}
\begin{equation}
\begin{split}
(p\oplus q)'_{\mu} = \left(1 -\frac12 \left(B -2A\right) \left(p'\cdot q' + q'^2 \right) \right) p'_{\mu} + q'_{\mu} + o(p'^3)
\end{split}
\end{equation}

Now it is clear that the expression found in \cite{BF} is the same as the one we found,  after a rescaling 
$(p,q) \to ((B -2A)p, (B -2A)q)$.

\section{Conclusions}
In this paper we tried to answer the question of ``how free'' we are to deform the crucial features of the momentum space in the relative locality framework: symmetry algebra, metric and composition law.
Thanks to the Nijenhuis-Richardson theorem the process of deforming the symmetry group is somewhat restricted to a change of coordinates on the space of generators, thus if we assume diffeomorphism invariance it is completely trivialized. Retaining the Lorentz symmetry group the problem translates in finding every possible covariant deformation of metric and composition law on momentum space, which is a much simpler problem than the general one. Having at hand all these deformations we went on to analyze consequences and features of some desirable conditions on the composition law, using perturbation expansions to avoid calculational obstructions.
In particular we found that at least at lowest order in perturbation theory the request of metric compatibility of the composition law implies a ``Snyder-like'' momentum space, the same reported in \cite{BF}. 
Some interesting phenomenological consequences can be extracted from these results: in particular the simplest one is that - given all the results reported - diffeomorphism invariance in momentum space implies that the first non-trivial order in the deformations is the second one, so looking for a first-order effect of the Planck scale would put on trial such a diffeomorphism invariance. The analysis we carried on here, like other ones in the usual relative locality framework, are classical, i.e. we have $\hbar=0$. It will be interesting to review the analysis brought here in the viewpoint of quantum mechanics, taking inspiration from works like \cite{Trev,Val1,Val2}.

\end{document}